*Article*

# Design and Implementation of an Automated Disaster-recovery System for a Kubernetes Cluster Using LSTM

**Ji-Beom Kim, Je-Bum Choi, and Eun-Sung Jung \***

Department of Software & Communications Engineering, Hongik University, Sejong 30016, Republic of Korea;
kjbeom@g.hongik.ac.kr, b989061@g.hongik.ac.kr
\* Correspondence: ejung@hongik.ac.kr

**Abstract:** With the increasing importance of data in the modern business environment, effective data management and protection strategies are gaining increasing research attention. Data protection in a cloud environment is crucial for safeguarding information assets and maintaining sustainable services. This study introduces a system structure that integrates Kubernetes management platforms with backup and restoration tools. This system is designed to immediately detect disasters and automatically recover applications from another kubernetes cluster. The experimental results show that this system executes the restoration process within 15 s without human intervention, enabling rapid recovery. This, in turn, significantly reduces the potential for delays and errors compared with manual recovery processes, thereby enhancing data management and recovery efficiency in cloud environments. Moreover, our research model predicts the CPU utilization of the cluster using Long Short-Term Memory (LSTM). The necessity of scheduling through this predict is made clearer through comparison with experiments without scheduling, demonstrating its ability to prevent performance degradation. This research highlights the efficiency and necessity of automatic recovery systems in cloud environments, setting a new direction for future research.

## 1. Introduction

In the modern era, data have become a crucial asset and a key to competitiveness for businesses. With the expansion of digitalization and cloud technology, data continues to increase in value. In this context, data loss poses a significant threat to businesses, highlighting the need for efficient data management and protection strategies. Data protection in cloud environments is vital, enabling companies to respond to various threats such as software errors, hardware failures, cyber-attacks, and natural disasters. The maintenance of data stability and reliability has emerged as a key element of business operations.

The importance of data backup is emphasized in this context. Data backup transcends mere information copying; it involves various functions, from fulfilling legal obligations to protecting against security threats, such as ransomware and hacking, and preparing for natural disasters [1-4]. Data backup strategies, such as the setting of recovery time objectives (RTOs) and recovery point objectives, play a crucial role in minimizing service disruptions and preventing data loss. These strategies are essential for organizations to protect their information assets from various risks and maintain sustainable services.

The significance of disaster recovery is an extension of data backup strategies. Disaster recovery provides more comprehensive security and stability in conjunction with data backup. Disaster-recovery solutions respond quickly and effectively to data loss or damage due to natural disasters, technical errors, and cyber-attacks. Such strategies ensure quick recovery and the provision of sustainable services, thus guaranteeing business continuity. Companies operating various applications and services in cloud environments must implement advanced disaster-recovery strategies.

Disaster Recovery as a Service (DRaaS) supports the quick recovery of data and applications in the event of a disaster [5]. This service is applied in various environments, and its importance has been increasingly highlighted in cloud usage. Cloud-based businesses and organizations require effective disaster-recovery strategies to ensure data and service continuity. DRaaS is designed to meet this requirement by considering the flexibility and scalability of cloud infrastructure to minimize business disruptions in the event of a disaster. DRaaS in cloud environments offers many advantages, including cost efficiency, fast recovery times, and accessibility, making it an essential element in modern business environments. Consequently, research on disaster recovery in cloud environments is being actively pursued [6-10].

Backup and restoration tools for Kubernetes are crucial for protecting the data and system state of container-based applications. These tools play an important role in rapidly restoring important data and configurations in the event of a failure [11]. Their usage supports the safe backup of critical data and configurations and quick restoration in case of failure or data loss.

The Kubernetes management platform provides an environment for the integrated management of various clusters and services. With the continuous increase in the complexity of cloud services and applications, consistent cluster management and deployment become essential. This platform enhances operational efficiency in complex environments, reduces operational burdens through resource optimization and automation, and emphasizes the need for a centralized management platform with the increase in the number of clusters. This reduces the possibility of errors and facilitates maintenance and monitoring. The Kubernetes management platform simplifies the manual management of individual clusters and supports consistent policy application, effective monitoring, and stable application deployment. This plays a significant role in enhancing the efficiency of the IT infrastructure for businesses and organizations, contributing to the achievement of business goals.

This paper presents a system that automatically recovers applications in a cluster. The system detects when a cluster loses functionality because of a disaster and automatically restores the services that were operating in the affected cluster to another cluster. Accordingly, even if a disaster occurs and the cluster loses functionality, automatic recovery is immediately initiated without the need for user intervention. Conventional disaster recovery involves delays due to human intervention [12], which can be minimized in an automated recovery, thus providing the advantage of faster recovery. Compared with manual recovery, an automated system responds immediately and quickly restores services to normal. In addition, it prevents mistakes that can occur during manual operations and ensures consistent recovery.

Our research has two technical contributions. First, it automates the entire recovery process, i.e., from event detection, including disasters and attacks, to querying backup files, selecting clusters for restoration, and executing restoration tasks. To automate all these functions, the system integrates Kubernetes management platforms with Kubernetes backup and restoration tools. Thus, user-intervention time is eliminated. This automation prevents mistakes that can occur during manual operations because of the complexity of the recovery process and ensures consistent recovery. In addition, automated recovery uses preallocated resources to perform tasks most optimally, thereby improving the system's overall performance and stability.

Second, when selecting a cluster for the restoration task, machine learning is used to predict the CPU usage required for selecting a cluster. While clusters can be selected using algorithms or rules, the time taken for restoration tasks must also be considered. Therefore, machine learning is used to predict the CPU usage of clusters to select the cluster for successful restoration. This minimizes service disruption with quick recovery times, positively affecting RTOs. Finally, the automated-recovery solution reduces management complexity and enhances the overall system stability by providing consistent recovery procedures across various systems or applications. This approach prevents mistakes that can occur during manual operations and ensures consistent recovery.

The remainder of this paper is structured as follows. Chapter 2 reviews current related research, highlighting the importance of recovery automation in cloud environments and presenting the unique aspects of our research. Chapter 3 describes the architecture of the proposed system, which integrates Kubernetes management platforms with backup and restoration tools. Chapter 4 introduces the data used in the experiments and explains the preprocessing methods used. Chapter 5 discusses the Long Short-Term Memory (LSTM) and how to train time-series data on the LSTM. Chapter 6 details the experimental environment and methods used. In Chapter 7, we present and analyze the experimental results. Finally, conclusions are provided in Chapter 8.

## 2. Related Work

Sousa et al. [13] explored two important concepts in cloud software engineering: "automatic recovery" and "job scheduling." Automatic recovery refers to the automatic restoration of services through an orchestration manager in the event of system failures, enhancing the reliability of cloud services and minimizing the response time to failures. Job scheduling involves the efficient allocation and scheduling of resources in the cloud environment, optimizing system performance, and reducing operational costs. These functionalities are essential for stable and efficient management of the cloud infrastructure, especially in environments requiring high availability and quick recovery times. The superiority of automatic recovery over manual recovery is demonstrated through experiments that simulate various failure scenarios and measure the reduction in the service-restoration time during automatic recovery.

In manual recovery, users must identify failures and decide on the type of system recovery, thus causing delays and errors. In contrast, automatic recovery precisely monitors the service status and attempts recovery automatically when the orchestration manager detects a failure, thus enhancing reliability. Container deployments, including automatic recovery, are considered part of the service-development process, restoring containers to a normal state after failures. The experimental results show that the orchestration manager continuously monitors the status of services and automatically restarts services upon detecting issues, thereby reducing service downtime, and enhancing system reliability, demonstrating that rapid service recovery is possible without manual intervention. This automatic-recovery mechanism proves crucial in the operation of cloud-based services.

Yu et al. [14] focused on the automatic recovery of applications within aerospace ground systems based on cloud computing. They emphasized developing and implementing recovery services to counter software errors. The core of the recovery service is the automatic-recovery capability of applications, which is aimed at improving stability and availability in the cloud computing environment. They explored the technical details related to software recovery strategies and provided experimental evaluations of the recovery time and capability. In addition, measures to automate application recovery concerning software failures in cloud computing environments were designed and implemented, including strategies for various recovery scenarios. Moreover, the efficiency and performance of these strategies were validated using real experiments. Yu at al. also presented an automated approach to maintain continuous access portals and ensure business continuity after application recovery, allowing users to use the service continuously without being aware of the recovery process. Their research, which focused on the automation of application recovery in cloud environments, presented a different approach from those of previous studies. While most research [15-17] has focused on data and system recovery, this study focused on automatic recovery at the application level, offering a new direction for enhancing the stability and availability of applications in cloud environments.

Previous studies have presented diverse approaches to automatic recovery in Kubernetes and cloud computing environments. Our research focuses on automatic recovery at the cluster level. In contrast to the research by Sousa et al. [13], which was focused on individual services or tasks, our approach involves the conducting of automatic recovery by targeting clusters, thus restoring applications from one cluster to another. Yu et al. [14] emphasized automated application recovery in aerospace ground systems based on cloud computing. This study proposes and implements automatic-recovery functions at the application level in response to software failures. It addresses application recovery strategies and experimentally evaluates recovery time and capability. Compared to the technique used in [14], which required standby server resources, our research enhances resource-utilization efficiency by conducting recovery in operational clusters. Yu et al. [14] addressed automatic recovery from functional loss of applications in the same environment, whereas we propose cluster-level automatic recovery, which automatically restores applications to a different cluster when the original cluster loses functionality.

## 3. Design

This section explains the structure of the automatic-recovery system, implemented by integrating Kubernetes backup and restoration tools with the Kubernetes management platform. This system automatically transfers the applications of a cluster to another cluster if the original cluster experiences a disaster and loses its connection to the Kubernetes management platform. The automatic recovery system is added to the cluster state monitoring part of the Kubernetes management platform and integrates by installing Kubernetes backup and restoration tools in the environment.

The proposed automatic-recovery system operates on the Kubernetes management platform and continuously monitors the state of a cluster. In the event of a disaster, the system is capable of monitoring and detecting events to identify the situation. Once a disaster is detected, the system performs resource comparison by comparing the resources of the affected cluster with those of other clusters to identify a cluster with superior resources. Subsequently, it verifies the name of that cluster and selects a target cluster for performing the restoration work based on the CPU usage of that cluster. Finally, the system executes the restoration process by using the backup files of the disaster-affected cluster in the selected cluster. This entire process comprises four main components, and Figure 1 shows the flowchart of the components of the proposed automatic recovery system. The structure and operation of each component are as follows.

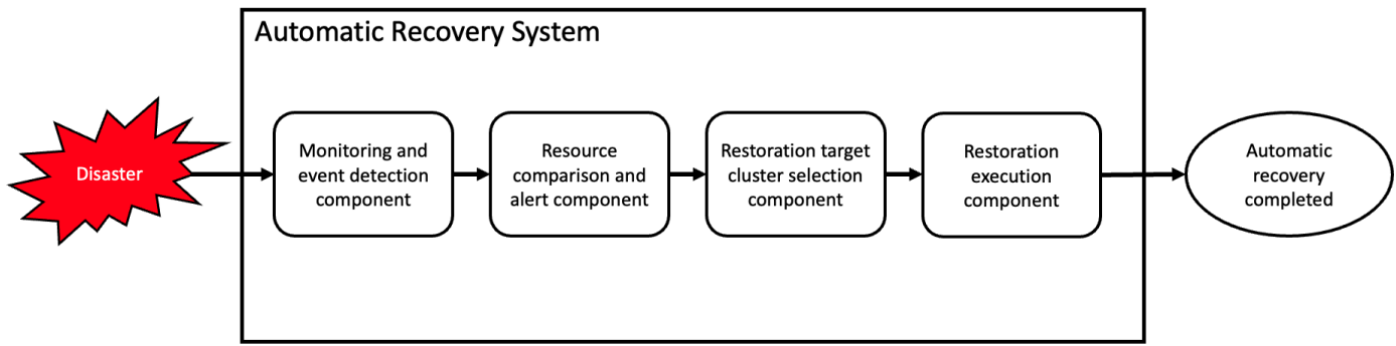

**Figure 1.** Flowchart of the Components of the Automatic Recovery System

- *Monitoring and event-detection* component: The Kubernetes management platform monitors the clusters it manages. In this study, we enhanced this feature to detect events when a managed cluster becomes disconnected and then transmit the name of that cluster to the resource-comparison component. Figure 2 shows the flowchart of the *monitoring and event-detection* component.

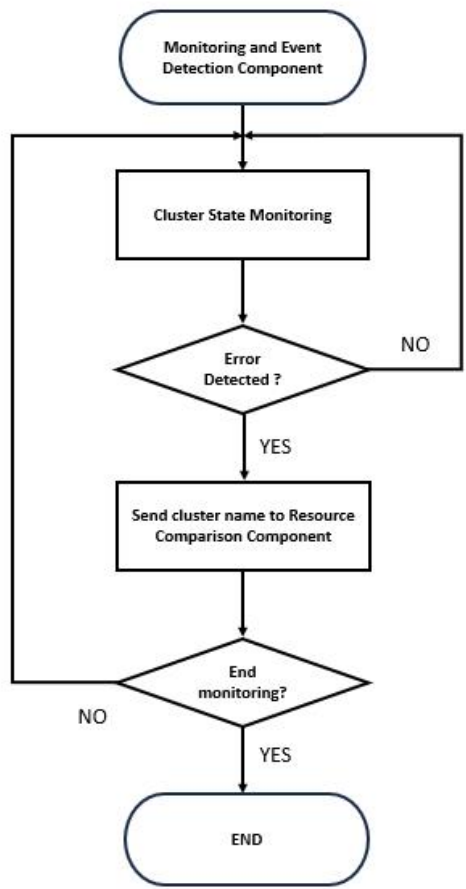

**Figure 2.** Flowchart of the monitoring and event-detection component

- *Resource-comparison-and-alert* component: By using the name of the disaster-affected cluster transmitted by the *monitoring and event-detection* component, the allocated CPU core counts of that cluster and the other managed clusters are compared to check if any cluster has more CPU cores than the disaster-affected cluster. If none of the other clusters match this criterion, the restoration procedure is halted and an alert is sent to the user, as proper restoration cannot be achieved. In contrast, if clusters are found with more CPU cores than those of the affected cluster, the names of such clusters are retrieved for the restoration process and sent to the *Restoration target cluster-selection* component. As there can be more than one cluster with more CPU cores than the affected cluster, multiple cluster names can be transmitted to the *Restoration target cluster-selection* component. Figure 3 shows the flowchart of the *Resource-comparison-and-alert* component.

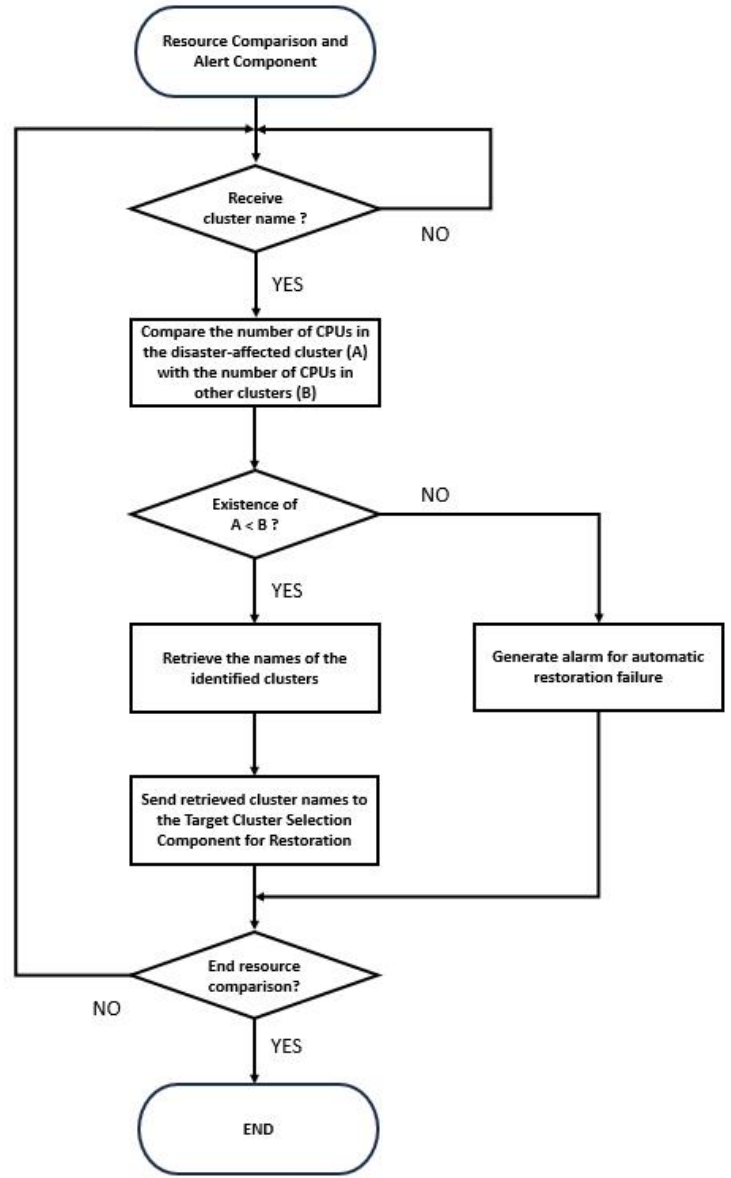

**Figure 3.** Flowchart of the Resource-comparison-and-alert component

- *Restoration target cluster-selection* component: The names of the clusters transmitted through the *Resource comparison* component are used to query their current CPU-utilization rates. The queried current CPU-utilization rates of these clusters are passed to a machine-learning model as parameters to predict CPU utilization, and the predicted CPU-utilization rates are returned. The cluster with the lowest predicted CPU-utilization rate among all predicted clusters is selected. Then, the name of the selected target cluster for restoration is transmitted to the *Restoration-execution* component. Figure 4 shows the flowchart of the *Restoration target cluster-selection* component.

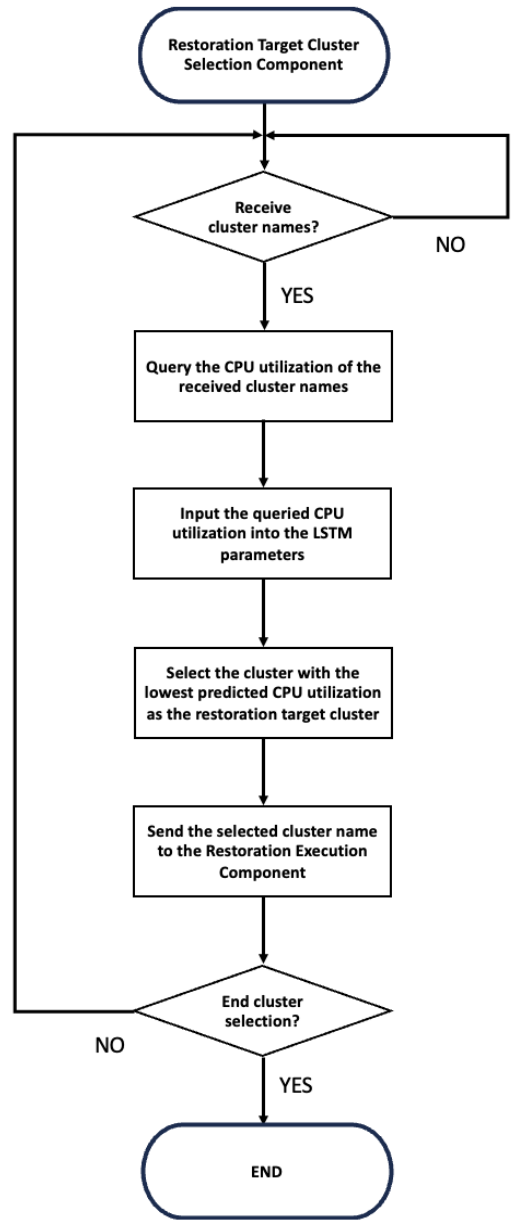

**Figure 4.** Flowchart of the Restoration target cluster-selection component

- *Restoration-execution* component: The cluster selected as the target for restoration and transmitted through the *Restoration target cluster-selection* component, executes the restoration command of the Kubernetes backup and restoration tool. This restoration command includes the location of the backup file of the disaster-affected cluster. When the restoration command is executed in the target cluster, the backup file is retrieved from its storage location and used to restore applications and other components. Figure 5 shows the flowchart of the *Restoration-execution* component.

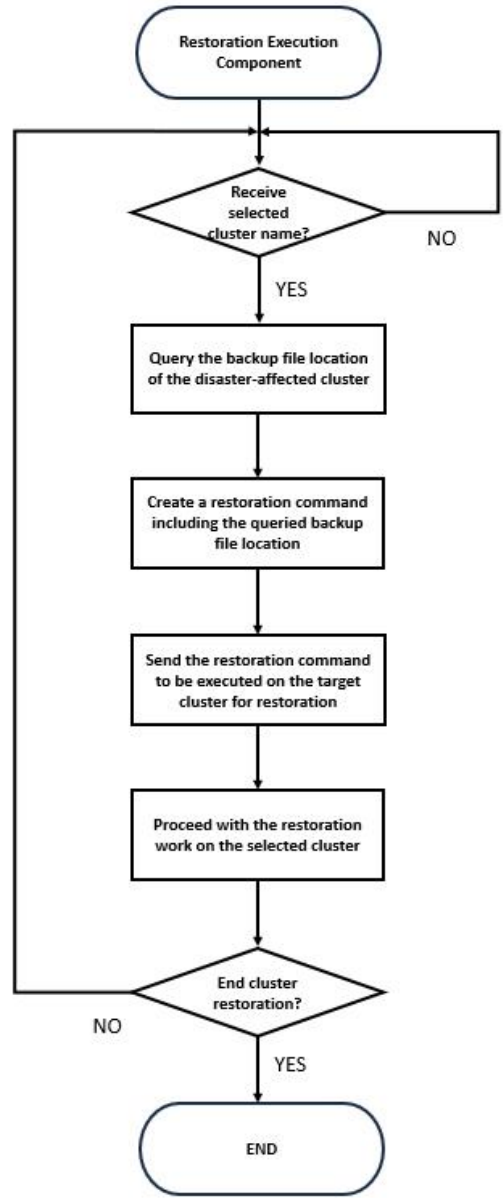

**Figure 5.** Flowchart of the Restoration-execution component

## 4. Experimental Data

*4.1. Google Cluster Trace Dataset*

We used the Google Cluster Trace dataset [18], specifically the ClusterData-2011_2 version, to train our cluster-state-prediction model. This dataset records the activity of a single cluster during May 2011, encompassing approximately 12,500 machines, 650,000 jobs, and 20 million tasks. The tasks mentioned herein refer to Linux programs executable on machines. The dataset includes detailed trace information about the behavior of jobs and tasks, resource allocation, and the activities of machines in the cluster. The dataset is categorized into various tables, including the Machine Event, Machine Attributes, Job Event, Task Event, Task Constraints, and Task Resource Usage Tables. Each table provides the following information:

- Machine Events Table: This table comprises one or more records of every machine in a cluster. Most of the records describe the machines present at the start of the trace. Event types include addition, removal, and update, and the CPU and memory capacities of each machine are standardized. The platform ID denotes the micro-architecture and chipset version of the machine; machines with the same ID can differ in terms of clock speed or number of cores.
- Machine Attributes Table: This table comprises key-value pairs representing the machine's characteristics, including kernel version, clock speed, and IP address. Values are expressed as strings if not in integer form, and "1" indicates a missing value.
- Job Events Table: This table includes the time, ID, type, user, and scheduling information of a job. Information about active (RUNNING) or pending (PENDING) jobs is also recorded, with each job containing scheduling constraints. The scheduling class contains the latency information of the job, and job names are provided as encrypted strings, repeated for multiple runs of the same program.
- Task Events Table: This table contains information such as timestamps, missing details, job ID, task index, machine ID, event type, username, scheduling class, priority, CPU cores, RAM, and local disk space requests. A task's priority is inversely proportional to its numerical value, with higher numbers indicating higher priority. "Free" denotes low priority, "production" is high priority, and "monitoring" is a priority for monitoring other low-priority tasks. Resource requests indicate the maximum CPU, memory, and disk space that a task can use, and exceeding these limits can restrict the task.
- Task Resource Usage Table: This table includes information such as the start and end times of the measurement period, job ID, task index, machine ID, CPU usage, memory usage, disk I/O time, and cache usage. It contains essential data for understanding the actual resource usage in the cluster, such as average CPU usage, normalized memory usage, average disk I/O time, and average local disk space usage.

These tables represent CPU-related resource usage data as normalized values. This normalization adjusts to a relative scale based on the maximum resource capacity of all tracked machines, with the maximum value standardized at 1.0. The CPU usage is measured in core-seconds per second; for instance, if a job consistently uses two cores, the usage rate is 2.0 core-seconds per second [18].

*4.2. Preprocessing*

According to Bi [19], the resource usage data in the Google Cluster Trace dataset is highly nonlinear, exhibiting erratic and highly variable characteristics. Figure 6 visualizes this by normalizing the CPU rate data by using the min–max normalization method and representing it in 5-min intervals, showcasing the irregular values of the CPU rate.

Like previous studies focusing on CPU utilization [19, 20], data preprocessing was initiated by extracting the start and end times as well as CPU-usage data from the *task resource usage* table. As the Google Cluster Trace dataset does not provide direct CPU use data for the cluster, the extracted data were aggregated. We then identified the earliest measurement start time and the latest measurement end time, creating 8352-time slots at 5-min intervals based on this range. Next, the CPU utilization for each time slot was aggregated according to the start and end times of each task. This preprocessed data was converted to a time series, batched at 5-min intervals, and then used to analyze trends in CPU utilization of the cluster.

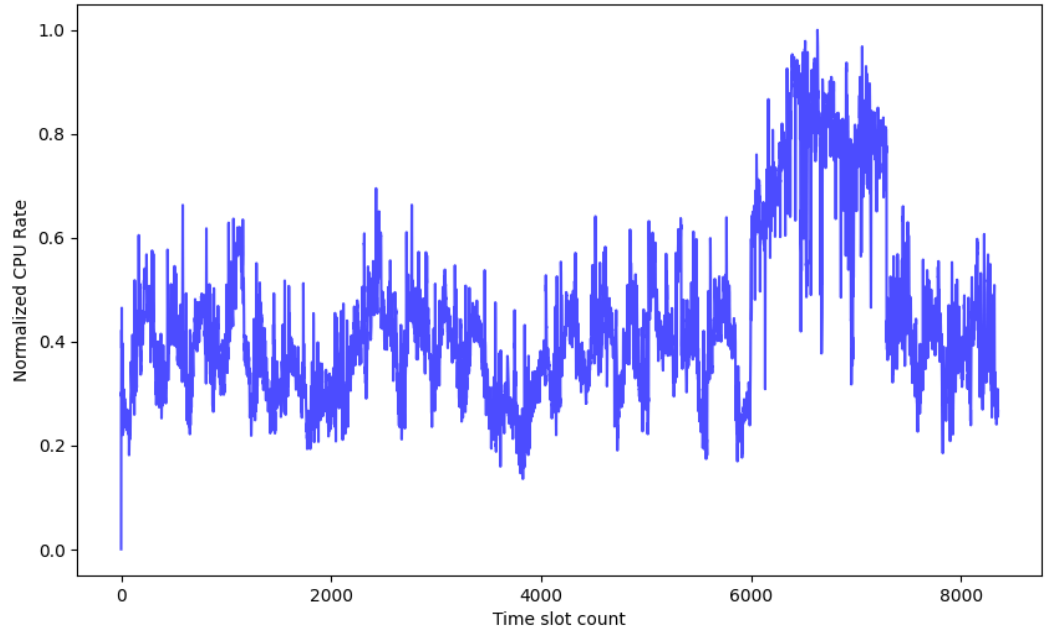

**Figure 6.** Time series of CPU usage

**5. LSTM**

LSTM [21] is an advanced recurrent neural network that effectively remembers the data sequences from previous time steps for future applications. The reason for choosing LSTM is that this model integrates gates and memory lines to effectively learn both long-term and short-term dependencies in the data. Due to these reasons, LSTM is commonly used in time series data prediction. In contrast to other models, LSTM excels in solving the gradient vanishing problem that occurs in long sequence data and has strengths in detecting and learning temporal dependencies. With these characteristics, LSTM demonstrates robust performance in capturing complex patterns and various time intervals in time series data.

*5.1. Sliding Window*

The lookback window represents the duration of past data provided to the model, with the current time as the reference point. The model is trained and performs predictions using data within this period. For example, if the lookback window is 24 hours, the model is trained and predicts based on data from the current time up to 24 hours ago.

The forecasting horizon refers to the time interval into the future that the model aims to predict, with the current time as the reference point. It determines how far into the future the model intends to make predictions. For instance, if the forecasting horizon is 6 hours, the model performs predictions for the timeframe from the current time to 6 hours ahead.

Applying the sliding window technique to chaotic long-term time series data, such as the Google Cluster Trace Dataset, offers several advantages:

- Nonlinear Behavior Detection: Chaotic data often exhibits distinct nonlinear dynamic patterns. Utilizing the sliding window allows the model to capture and learn these nonlinear patterns.
- Temporal Dependencies: Chaotic time series data involves crucial temporal dependencies. Sliding window considers data within specific periods, leading to a more accurate understanding of temporal dependencies.
- Adaptability and Model Generalization: Sliding window aids the model in adapting to the dynamic nature of the data. Chaotic data can be challenging for prediction, but through the sliding window, the model can learn and generalize patterns within given periods.

We explored the optimal lookback window and forecasting horizon through experiments on the Google Cluster Trace Dataset. In conclusion, the best prediction performance was observed with a lookback window of 3 and a forecasting horizon of 1.

*5.2. Prediction Model Architecture*

Figure 7(a) illustrates the process of the LSTM model learning time series data. It depicts the process of constructing and predicting time series data using the LSTM model. Initially, the time series data is normalized to the [0,1] range through Min-Max scaling. Subsequently, the data is divided into X values and Y values by setting the Lookback Window and Forecasting Horizon of the sliding window. Here, X values correspond to data from past times, while Y values correspond to data for future times. The data is then split into training and testing sets through Data Split, with a split ratio of 0.2. The LSTM sequentially receives X values from the training data, and Y values are used as labels in supervised learning. The generated LSTM model has a total of 198,273 trainable parameters, and each model undergoes training for 50 epochs.

Figure 7(b) illustrates the process of using the trained model to make predictions. The X values from the Test Data are sequentially input into the trained LSTM model, and the model predicts the Y values corresponding to the X values through the prediction process, representing the CPU rate. As these predictions are normalized, they undergo a Denormalization process to convert them back into actual prediction values.

In terms of performance evaluation, our LSTM model demonstrates a performance with MAE (Mean Absolute Error) = 0.0278, MAPE (Mean Absolute Percentage Error) = 3.5268, and $R^2$ (Coefficient of Determination) = 0.9598.

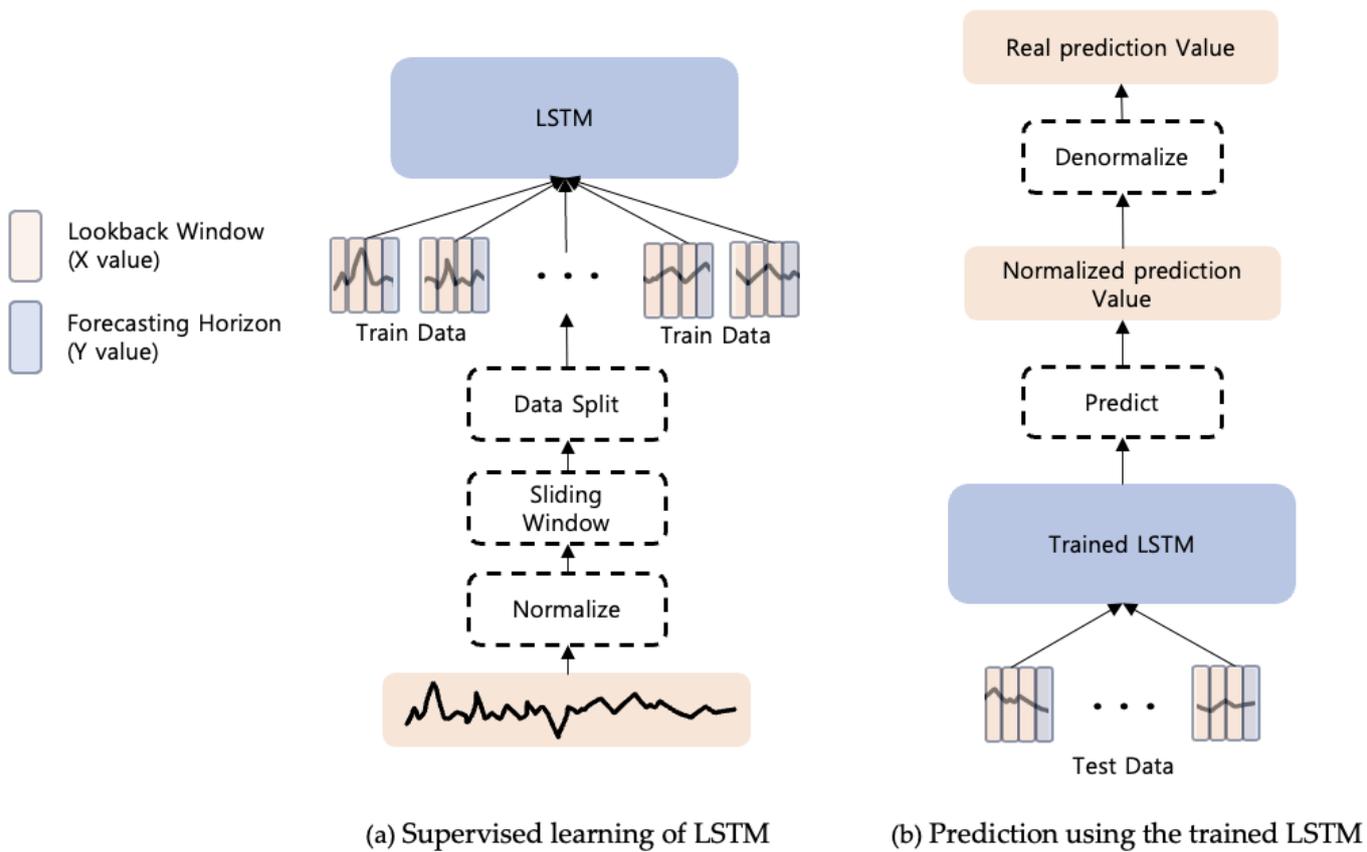

**Figure 7.** LSTM workflow. (a) Supervised learning of LSTM with normalization, sliding window, data split; (b) Prediction using the trained LSTM with denormalization.

Therefore, in our research, we apply the LSTM model to the scheduling of time series data. This model effectively detects nonlinear behavior patterns, understands temporal dependencies, and adapts to chaotic long-term time series data through adaptability and model generalization.

## 6. Experiment

In this study, we conducted two experiments. The first is an experiment on the automatic recovery of a cluster, and the second is a scheduling experiment that uses LSTM. Both experiments were conducted in a cloud environment. The primary objective of this study was to demonstrate the efficiency and performance of Kubernetes cluster automatic recovery according to the recovery time and prove the necessity of scheduling using machine learning. The experimental environment uses AWS EC2 instances and S3 buckets, with the EC2 instances running Ubuntu Server 22.04 LTS (HVM), SSD Volume Type. The open-source tools, Rancher and Velero, are used in the Kubernetes management platform and Kubernetes backup and restoration tools, respectively. The proposed automatic recovery system is implemented by adding an automatic recovery feature to the Rancher source code. The Rancher used in the experiment uses Rancher that includes the automatic recovery system.

*6.1. Automated Cluster Recovery*

Figure 8 shows the Rancher configuration for the automatic cluster-recovery experiment. The Kubernetes cluster running Rancher is referred to as the Rancher cluster. Although Rancher can operate as a container on a single server, in our experiment, it runs within a Kubernetes cluster because of the use of Velero. Clusters 1 and 2, linked to the Rancher cluster, each comprise one master node and two worker nodes. The master node is a t2.medium EC2 instance type with 2 vCPU, 4 GB memory, and 30 GB storage, and the worker nodes use t2.small EC2 instances with 1 vCPU, 2 GB memory, and 30 GB storage.

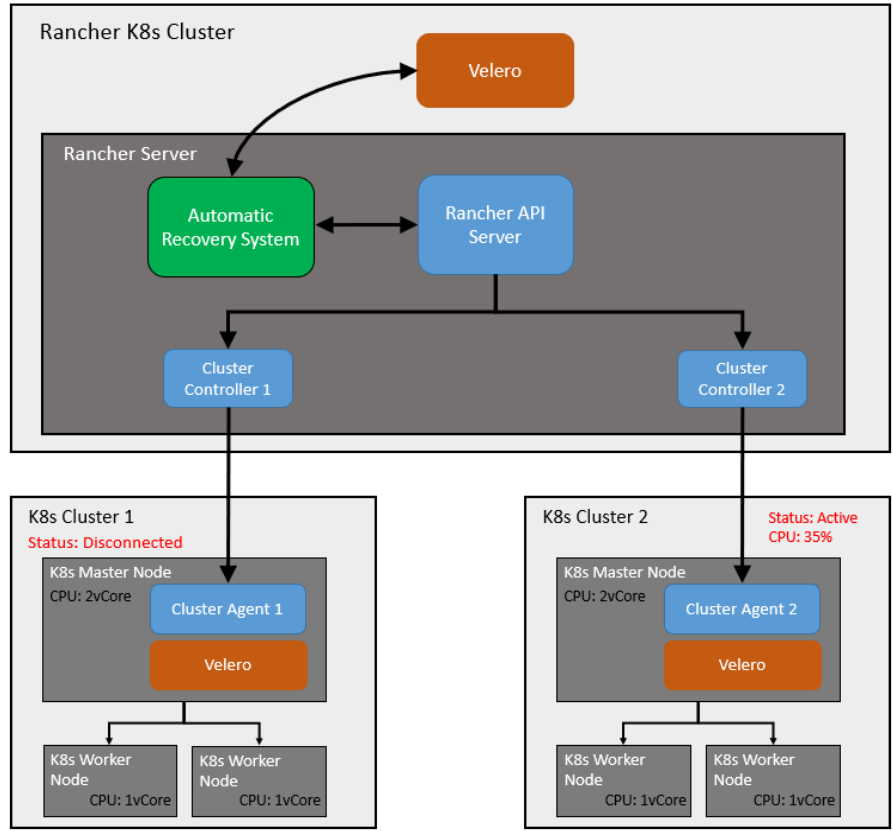

**Figure 8.** Architecture for the automated cluster-recovery experiment

Figure 9 shows the flowchart of the cluster auto-recovery experiment. Velero, connected to the AWS S3 bucket, is installed in all Rancher Cluster, Cluster 1, and Cluster 2. In step 2, if Cluster 1 loses functionality due to a disaster, Rancher detects this. In step 3, the detected status of Cluster 1 changes to disconnected, triggering the auto-recovery system. In step 4, the auto-recovery system uses Velero to look up the latest backup file of Cluster 1. In step 5, the restore command, including the name of this backup file, is transmitted to Cluster 2. In step 6, Cluster 2 executes the received command, using the backup file in the S3 bucket to restore the application that was running in Cluster 1 to Cluster 2.

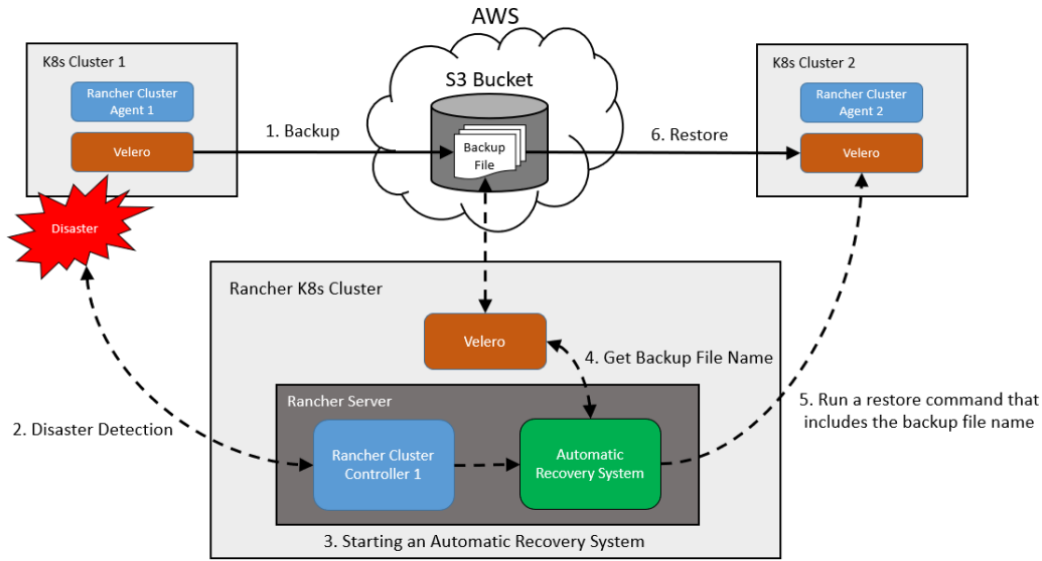

**Figure 9.** Flowchart of the automated cluster-recovery experiment

The automatic-recovery experiment was conducted as follows. First, Nginx was run on Cluster 1, and a backup file was created in the AWS S3 bucket by using the Velero backup command. Then, the master node of Cluster 1 was stopped to disconnect it from the Rancher cluster. When the disconnection is detected by the Rancher cluster, the proposed automatic-recovery system is initiated for the cluster. Because Clusters 1 and 2 both have a total of 4vCPU allocated, Cluster 2 is a viable cluster for restoration. In this experiment, because Cluster 2 was the only target cluster available for restoration, the restoration proceeded to Cluster 2 without scheduling. First, Velero in the Rancher cluster queried the latest Nginx backup file in the S3 bucket and sent the restoration command, including the name of the backup file, to Cluster 2. Cluster 2 executed the received command, restoring the Nginx running on Cluster 1 by using the backup file in the S3 bucket.

In this experiment, excluding the process of creating the backup file, the time taken from when the master node of Cluster 1 was stopped, to the creation of an artificial disaster scenario, to when the Nginx on Cluster 2 was completely restored was measured. This experiment was repeated 10 times to measure the time. The time taken from executing the Velero restoration command in Cluster 2 to the completion of the restoration is referred to as the restoration time. The time utilized by the process, excluding the restoration time, is attributed to the proposed system. The measurement and analysis of these times are explained in Section 7.

*6.2. LSTM-based Scheduling*

This experiment is based on LSTM-based scheduling, utilizing a model trained on Google Cluster Trace data. The data includes cluster CPU usage rates at 5-minute intervals, enabling the training of a model to predict the cluster CPU usage rate 5 minutes into the future. The model, built on insights from five experiments, analyzes CPU usage rates over the past 15 minutes to predict the rate for the next 5 minutes. The choice of predicting the CPU usage rate for the next 5 minutes is due to the high volatility in the Google Cluster Trace data. Through LSTM-based scheduling, the algorithm selects clusters with higher stability compared to algorithms considering only the current state. In the experiments, Google Cluster Trace data is applied to predict CPU usage rates in the current environment. The cluster's CPU usage rates are adjusted at 5-minute intervals, aligned with the patterns in the Google Cluster Trace data. This process aims to validate the model's prediction accuracy by adapting it to the real environment. In the event of a disaster, the model mimics data patterns, predicts CPU usage based on applied data in the actual cluster environment, and selects the cluster with the lowest predicted CPU usage rate for recovery.

Figure 10 shows the Rancher configuration for the LSTM-based scheduling experiment. Each of the six clusters linked to the Rancher cluster comprises one master node and two worker nodes. The master node is a t2.medium EC2 instance with 2 vCPU, 4 GiB memory, and 30 GiB storage, whereas the worker nodes are t2.small EC2 instances with 1v CPU, 2 GiB memory, and 30 GiB storage. The CPU-utilization rates of the clusters were varied, which was intentional for the scheduling test.

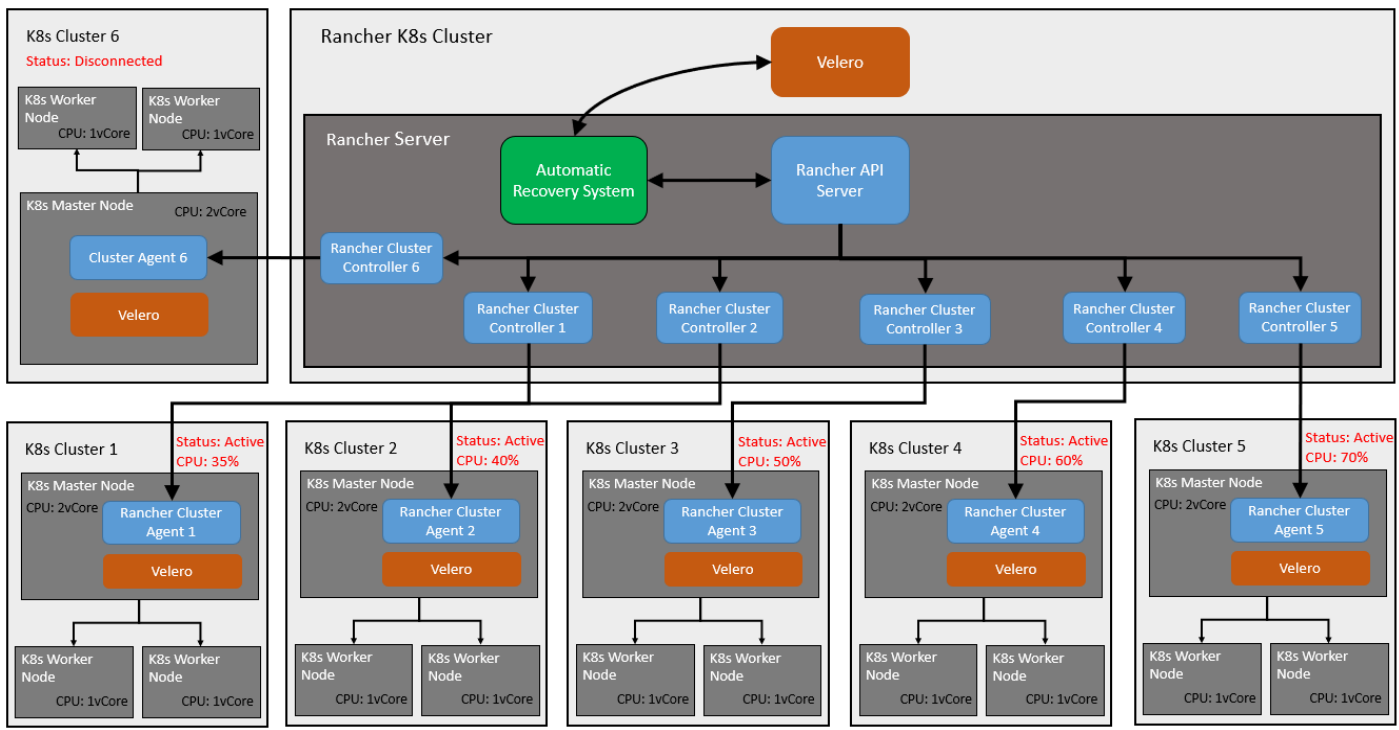

**Figure 10.** Architecture used in the LSTM-based scheduling experiment

Figure 11 illustrates a YAML file for a pod operating to adjust the CPU-utilization rates of the clusters. This pod uses a very resource-light busybox image. The "requests" and "limits" represent the CPU usage request and maximum limit, respectively. By adjusting these values and running them, the CPU-utilization rates of each cluster can be manipulated. In the example, "requests" and "limits" were set to 200 m to achieve 5% utilization of the total 4vCPU of the cluster.

```yaml
apiVersion: v1
kind: Pod
metadata:
  name: dummy-pod
  namespace: dummy
spec:
  containers:
  - name: busybox
    image: busybox
    command: ["sh", "-c", "while true; do sleep 1; done"]
    resources:
      requests:
        cpu: "200m"
      limits:
        cpu: "200m"
```

**Figure 11.** Dummy application YAML

Figure 12 shows the flowchart of the LSTM-based scheduling experiment. Velero, connected to the AWS S3 bucket, is installed in all clusters. In step 1, Cluster 6 uses Velero to create a backup file of the application running in it in the S3 bucket. In step 2, if Cluster 6 loses functionality due to a disaster, Rancher detects this. In step 3, the detected status of Cluster 6 changes to disconnected, triggering the auto-recovery system. In step 4, the auto-recovery system uses Velero to look up the latest backup file of Cluster 6. In step 5, the restore command, including the name of this backup file, is transmitted to the target cluster selected through LSTM-based scheduling. In the figure, it is transmitted to Cluster 1, which has the lowest predicted CPU usage. In step 6, the target cluster executes the received command, using the backup file in the S3 bucket to restore the application that was running in Cluster 6 to the target cluster.

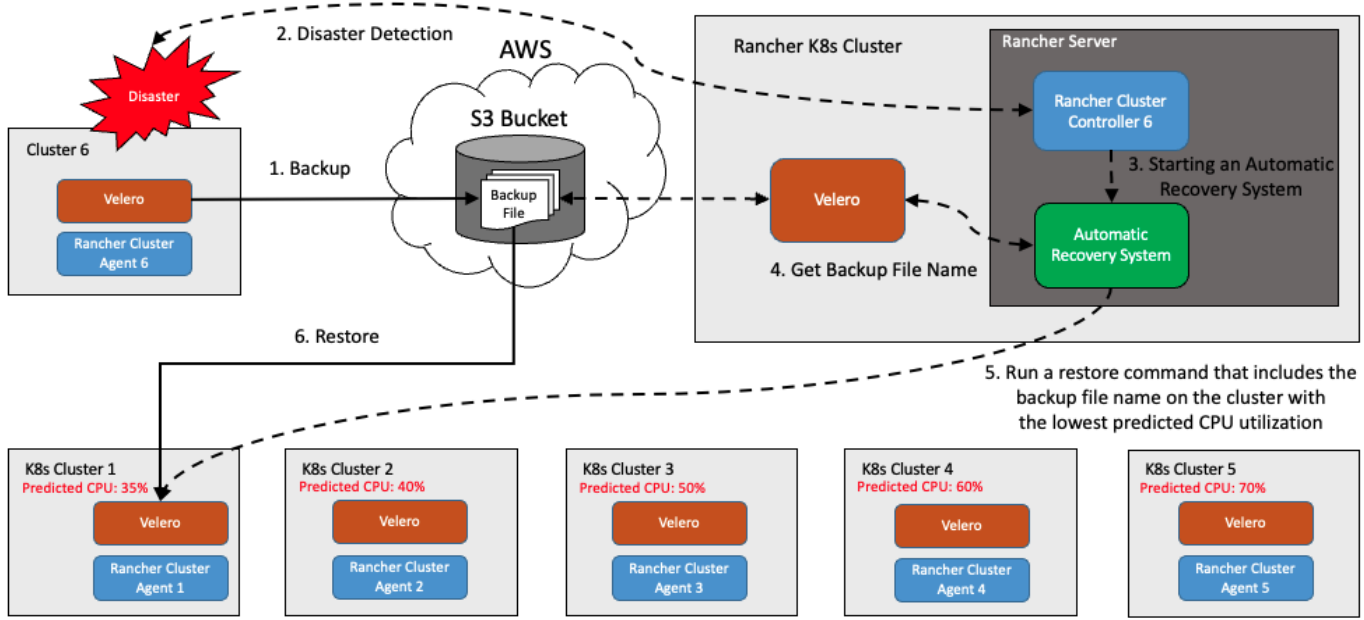

**Figure 12.** Flowchart of the LSTM-based scheduling experiment

The procedure of the LSTM-based scheduling experiment is as follows. First, a dummy application is run on Cluster 6. This running application is backed up to the AWS S3 bucket by using the *Velero backup* command. Then, the master node of Cluster 6 is stopped to disconnect it from the Rancher cluster. After the Rancher cluster detects the disconnection, our automatic-recovery system is initiated for the cluster. As Cluster 6 contains a total of 4vCPU, and Clusters 1, 2, 3, 4, and 5 each have 4vCPU allocated, they are all potential restoration targets. In this experimental environment, multiple target clusters are available for restoration, and thus scheduling is conducted. The current CPU-utilization rates of the clusters are used as input values for the LSTM-based CPU prediction. The cluster with the lowest predicted CPU-utilization rate is selected as the target cluster. Then, Velero in the Rancher cluster queries the latest backup file of the dummy application in the S3 bucket and sends the restoration command, including the name of the backup file, to the target cluster. The target cluster executes the received command, restoring the dummy application running on Cluster 6 by using the backup file in the S3 bucket. In our study, this experiment was conducted 10 times, measuring the CPU-utilization rates of the clusters each time. Additionally, an experiment in which the target cluster was randomly selected instead of using LSTM-based scheduling was conducted 10 times in the same manner. These two experiments are compared and analyzed in Section 7.

## 7. Results and Discussion

### 7.1. Automated Cluster Recovery

In Table 1, A represents the time taken to restore Nginx in our environment, B is the time taken for automatic recovery, and A–B is the time excluding the restoration time from the total time spent on automatic recovery. The restoration of Nginx in our environment takes 20 s, which is the time required for the restoration operation to be executed. In the automatic-recovery experiment, restoration was completed in an average of 27 s across 10 trials, within a range of 20–34 s. Excluding the time taken for the restoration operation, an additional 0–14 s was added. This time is attributed to the delay caused by Rancher's 15-s interval for detecting cluster disconnections. However, even if a user were to perform manual recovery, the inevitable delay would occur only after a Rancher detects cluster disconnection. Excluding this inevitable delay, the actual additional delay caused by the automatic-recovery operation is determined to be less than 1 s. Therefore, the experiment proves the efficiency of automatic recovery by eliminating the delay that is caused by user intervention in a manual recovery process.

**Table 1.** Results of the time taken to recover Nginx

| Case | Recovery Time (A) | Restoration Time (B) | (A–B) |
|---|---|---|---|
| 1 | 23 | 20 | 3 |
| 2 | 34 | 20 | 14 |
| 3 | 28 | 20 | 8 |
| 4 | 26 | 20 | 6 |
| 5 | 32 | 20 | 12 |
| 6 | 22 | 20 | 2 |
| 7 | 23 | 20 | 3 |
| 8 | 25 | 20 | 5 |
| 9 | 30 | 20 | 10 |
| 10 | 27 | 20 | 7 |
| AVG | 27 | 20 | 7 |

*7.2. LSTM-Based Scheduling*

In this section, the importance of CPU-utilization prediction is explored in the automated-recovery process. First, an experiment selecting the target cluster for restoration based on CPU-utilization prediction by using LSTM is compared with an experiment randomly selecting the target cluster. In each experiment, restoration operations were conducted 10 times on five clusters, starting from the same initial state, and the variations in CPU utilization in each cluster were analyzed. According to Gusev et al. [23], when CPU utilization exceeds 80%, the occurrence of system performance degradation is highly possible. Thus, maintaining a stable level of CPU utilization in the cluster is important. This study thoroughly analyzed the results of both experiments, presenting a comparative analysis.

Table 2 shows the results of randomly selecting a target cluster among five clusters and performing restoration 10 times. The first restoration occurred in Cluster 5, demonstrating the highest initial CPU utilization, followed by several other restorations. After 10 restoration operations, Clusters 4 and 5 reached 85% CPU utilization, indicating a high probability of performance degradation. Moreover, CPU utilization across clusters demonstrates significant variation.

**Table 2.** Results of randomly scheduling clusters

| Restoration Count | Cluster 1 | Cluster 2 | Cluster 3 | Cluster 4 | Cluster 5 |
|---|---|---|---|---|---|
| Initial State | 35% | 40% | 50% | 60% | 70% |
| 1 | 35% | 40% | 50% | 60% | 75% |
| 2 | 40% | 40% | 50% | 60% | 75% |
| 3 | 40% | 40% | 50% | 65% | 75% |
| 4 | 40% | 40% | 50% | 70% | 75% |
| 5 | 40% | 40% | 50% | 70% | 80% |
| 6 | 40% | 40% | 50% | 70% | 85% |
| 7 | 40% | 40% | 55% | 70% | 85% |
| 8 | 40% | 40% | 55% | 75% | 85% |
| 9 | 40% | 40% | 55% | 80% | 85% |
| 10 | 40% | 40% | 55% | 85% | 85% |
| Clusters with CPU utilization under 80% | O | O | O | X | X |

Table 3 presents the results of selecting the target cluster for restoration based on CPU-utilization prediction with LSTM and conducting 10 restorations. The first restoration began in Cluster 1, which has the lowest initial CPU-utilization rate, and subsequent restoration operations targeted clusters with the lowest rate of CPU utilization at each point. After completing 10 restoration operations, none of the clusters exceeded 80% CPU utilization. Moreover, all clusters maintained a stable state, with CPU utilization between 55% and 70%.

**Table 3.** Results of scheduling clusters based on CPU-usage prediction by using LSTM

| Restoration Count | Cluster 1 | Cluster 2 | Cluster 3 | Cluster 4 | Cluster 5 |
|---|---|---|---|---|---|
| Initial State | 35% | 40% | 50% | 60% | 70% |
| 1 | 40% | 40% | 50% | 60% | 70% |
| 2 | 45% | 40% | 50% | 60% | 70% |
| 3 | 45% | 45% | 50% | 60% | 70% |
| 4 | 50% | 45% | 50% | 60% | 70% |
| 5 | 50% | 50% | 50% | 60% | 70% |
| 6 | 55% | 50% | 50% | 60% | 70% |
| 7 | 55% | 55% | 50% | 60% | 70% |
| 8 | 55% | 55% | 55% | 60% | 70% |
| 9 | 60% | 55% | 55% | 60% | 70% |
| 10 | 60% | 60% | 55% | 60% | 70% |
| Clusters with CPU utilization under 80% | O | O | O | O | O |

The results of these two experiments demonstrate the importance of scheduling based on CPU-utilization prediction. As shown in Table 2, in Experiment 1, restoration was started in Cluster 5, which had high initial CPU utilization, leading to performance degradation. Eventually, CPU utilization in Clusters 4 and 5 reached 85%, increasing the risk of performance degradation and showing an imbalance in utilization among clusters. In contrast, in Experiment 2, (Table 3) restoration was started in clusters with lower CPU utilization and LSTM for prediction, which helped maintain system balance. All clusters maintained a stable state, not exceeding 80% CPU utilization, and a utilization rate between 55% and 70% indicates efficient resource usage.

This comparison proves that prediction-based scheduling by using LSTM is crucial for resource management and performance degradation prevention in cloud environments. When the prediction model was used to determine the order of restoration operations, it enhanced the overall system performance, increased resource usage efficiency, and minimized the risk of potential performance degradation.

## 8. Conclusions

This study focused on the efficiency of a Kubernetes cluster automatic-recovery system and the effectiveness of scheduling methods using LSTMs. Compared with previous studies that mainly explored automatic recovery and restoration at the service or application level, this study experimentally verified the feasibility and effectiveness of automatic recovery by targeting the entire cluster. The contributions of this study are two-fold: the reduction in recovery time and the efficient utilization of resources. In addition, the scheduling method based on CPU-utilization prediction by using LSTM can optimize the selection of the target cluster for restoration, contributing to the overall system performance and stability.

Consequently, this research significantly contributes to the design and operation of automatic-recovery systems in cloud environments, opening new horizons for future research. In addition, we emphasize the need for further research on the applicability of automatic recovery in various cloud environments and system optimization by integrating AI technologies. This could stimulate research and innovation to advance cloud computing.